\documentclass[apj, iop]{emulateapj}

\usepackage{amsmath,amssymb}
\usepackage{graphicx}
\usepackage{natbib}
\def\Omm{{\Omega_m}}
\def\Ommz{{\Omega_m^{\,z}}}

\def\Omk{{\Omega_k}}
\def\Oml{{\Omega_{\Lambda}}}

\def\aap{A\&A}
\def\apj{ApJ}

\def\apjl{ApJ}
\def\mnras{MNRAS}
\def\araa{ARA\&A}

\def\prd{Phys. Rev. D}
\def\nat{Nat}

\newcommand{\beq}{
\begin{equation}
}
\newcommand{\eeq}{
\end{equation}
}
\newcommand{\kms}{\,{\rm km\,s^{-1}}}
\newcommand{\msun}{\ensuremath{{{\rm M}}_{\scriptscriptstyle \odot}}}

\def\simlt{\mathrel{\rlap{\lower 3pt\hbox{$\sim$}}\raise 2.0pt\hbox{$<$}}}
\def\simgt{\mathrel{\rlap{\lower 3pt\hbox{$\sim$}} \raise 2.0pt\hbox{$>$}}}
\newcommand{\vta}[1]{\vert \boldsymbol{a}_{#1}        \vert}

\newcommand{\vtl}   {\vert \boldsymbol{       {\ell}} \vert}

\def\spose#1{\hbox to 0pt{#1\hss}}
\newcommand{\lta}{\mathrel{\spose{\lower 3pt\hbox{$\mathchar"218$}}
      \raise 2.0pt\hbox{$\mathchar"13C$}}}
\newcommand{\gta}{\mathrel{\spose{\lower 3pt\hbox{$\mathchar"218$}}
      \raise 2.0pt\hbox{$\mathchar"13E$}}}

\def\gsim{ \lower .75ex \hbox{$\sim$} \llap{\raise .27ex \hbox{$>$}} }
\def\lsim{ \lower .75ex\hbox{$\sim$} \llap{\raise .27ex \hbox{$<$}} }

\def\kmps{{\rm\thinspace km \thinspace s^{-1}}}

\def\beq{\begin{equation}}
\def\eeq{\end{equation}}

\def\Omm{{\Omega_m}}
\def\Ommz{{\Omega_m^{\,z}}}

\def\Omk{{\Omega_k}}
\def\Oml{{\Omega_{\Lambda}}}

\shorttitle{Accretion and spin}
\shortauthors{Volonteri et al.}

\begin{document}


\title{The evolution of active galactic nuclei and their spins}


\author{M. Volonteri\altaffilmark{1,2}, M. Sikora\altaffilmark{3}, J.-P. Lasota\altaffilmark{1,4} and A. Merloni\altaffilmark{5}}


\altaffiltext{1}{Institut d'Astrophysique de Paris, 98bis Bd. Arago, 75014, Paris, France}
\altaffiltext{2}{Astronomy Department, University of Michigan, 500 Church St. , Ann Arbor, MI 48109}
\altaffiltext{3}{Nicolaus Copernicus Astronomical Center, Bartycka 18, 00-716 Warszawa, Poland}
\altaffiltext{4}{Astronomical Observatory, Jagiellonian University, ul. Orla 171, 30-244 Krak\'ow, Poland}
\altaffiltext{5}{Max-Planck-Institut f\"ur Extraterrestrische Physik, Giessenbachstr., D-85741, Garching, Germany}


\begin{abstract}
Massive black holes (MBHs) in contrast to stellar mass black holes are expected to substantially change their properties over their lifetime. MBH masses increase by several order of magnitude over the Hubble time, as illustrated by {So{\l}tan's} argument. MBH spins also must evolve through the series of accretion and mergers events that grow the MBH's masses. We present a simple model that traces the joint evolution of MBH masses and spins across cosmic time. Our model includes MBH-MBH mergers, merger-driven gas accretion, stochastic fueling of MBHs through molecular cloud capture, and a basic implementation of accretion of recycled gas. This approach aims at improving the modeling of low-redshift MBHs and AGN, whose properties can be more easily estimated observationally. Despite the simplicity of the model, it captures well the global evolution of the MBH population from $z\sim 6$ to today. Under our assumptions, we find that the typical spin and radiative efficiency of MBHs decrease with cosmic time because of the increased incidence of stochastic processes in gas-rich galaxies and MBH-MBH mergers in gas-poor galaxies. At $z=0$ the spin distribution in gas-poor galaxies peaks at spins $0.4-0.8$, and  is not strongly mass dependent. MBHs in gas-rich galaxies have a more complex evolution, with low-mass MBHs at low redshift having low spins, and spins increasing at larger masses and redshifts. We also find that at $z>1$  MBH spins are on average highest in high luminosity AGN, while at lower redshifts these differences disappear.

\end{abstract}

\keywords{Black hole physics --- Galaxies: active, nuclei}

\section{Introduction}
Astrophysical black holes span a large  range of masses, from the remnants of stellar evolution to monsters weighing by themselves almost as much as a dwarf galaxy. Notwithstanding the several orders of magnitude difference between the smallest and the largest black hole known, all of them can be described by only two parameters: mass and spin. So, besides their masses, $M$, astrophysical black holes are completely characterized by their dimensionless spin parameter, $a \equiv J_h/J_{max}=c \, J_h/G \, M^2$, where $J_h$ is the angular momentum of the black hole, and $0\le a\le 1$.  

Many theoretical efforts have focused on the mass growth of massive black holes (MBHs) and on their feedback onto the host \citep[see, e.g.,][]{haehnelt1993, haiman2000,kauffmann2000,wyithe2003,VHM,Hopkins2006,Croton2006}.   Spin has received less attention in the cosmological context \citep{Moderski1998,Volonterietal2005,2005ApJ...620...59S,LPC09,Fanidakis11,Barausse12}, but it has great relevance for the overall growth of MBHs, as follows.

First, geometrically thin and optically thick  accretion disks radiate with an efficiency, $\epsilon_{\rm rad}$, which is almost equal to the mass-to-energy conversion efficiency, $\epsilon_{\rm rad}\simeq\epsilon \equiv 1- E_{\rm ISCO}$, where $E_{\rm ISCO}=\sqrt{1-2/(3r_{\rm ISCO})}$ is the specific energy of the gas particle (in rest mass-energy units) in the innermost stable ($=$ marginally stable) circular orbit (ISCO), and $r_{\rm ISCO}$  is the radius of this orbit in $GM/c^2$ units. This radius and therefore radiation efficiency  depend solely on the BH spin parameter $a$. Maximal efficiency ($\epsilon \simeq 0.42$) is achievable by disks rotating around maximally spinning BHs; it drops to $\simeq 0.06$ for non-spinning BHs, and to $\simeq 0.05$ for maximally counter-rotating BHs. This entails a dependence of the  BH mass-growth rate on the spin value, implying longest growth time scales for larger positive spins.  More precisely, for a hole accreting at the Eddington rate, the black hole mass increases with time as:

\beq
M(t)=M(0)\,\exp\left({{1-\epsilon}\over{\epsilon}} \frac{t}{t_{\rm Edd}}\right), 
\eeq

where $t_{\rm Edd}=M_{\rm BH} c^2/L_{\rm Edd}=\frac{\sigma_T \,c}{4\pi \,G\,m_p}=0.45\,{\rm Gyr}$. The higher the spin, the higher $\epsilon$, implying longer timescales to grow the MBH mass by the same number of e--foldings.

The radiative efficiency is also the fundamental free parameter in the {So{\l}tan argument \citep{Soltan1982} and, more recently, in synthesis models \citep[e.g.,][]{Merloni08}  which relate the integrated MBH mass density to the integrated emissivity of  the AGN population, via the integral of the luminosity function  of quasars. If the average efficiency of converting  accreted mass into luminosity is $\epsilon=L/\dot{M}c^2$, then the MBH will increase its mass by $\dot{M}=(1-\epsilon)\dot{M}_{in}$, accounting for the fraction of the inflowing mass, $\dot{M}_{in}$, that is radiated away. Applying this argument to the whole MBH population,  the MBH mass density, $\rho_{BH}$, can be related to the integral of the luminosity function of quasar, $\Psi(L,z)$, with the radiative efficiency being a free parameter:

\beq
\rho_{BH}(z)=\int_z^\infty\frac{d t}{d z}d z
\int_0^\infty \frac{(1-\epsilon)\,L}{\epsilon c^2}\Psi(L,z)d L.
\eeq

Recent results suggest that this approach might be too simplistic, as the radiative efficiency evolves along the cosmic time. \cite{Wang2009} for instance suggests that quasars at the peak of their activity  ($z\sim2$) have high radiative efficiencies, hence  high spins. At later times ($z<1$), however,  the average radiative efficiency  decreases, hinting to lower spins.  This paper addresses  these issues, by providing spin distributions of MBHs as a function of cosmic epoch. 

MBH spins also affect the incidence of MBHs in galaxies, via the ``gravitational recoil" mechanism. When the members of a black hole binary coalesce, the center of mass of the coalescing system recoils due to the non-zero net linear momentum carried away by gravitational waves in the coalescence. If this recoil were sufficiently violent, the merged hole would break loose from the host and leave an empty nest.  Recent breakthroughs in numerical relativity have allowed reliable computations of black hole mergers and recoil velocities, taking the effects of spin into account. Non-spinning  MBHs, or binaries where MBH spins are {\it aligned} with the orbital angular momentum are expected to recoil with velocities below 200 $\rm{km\,s^{-1}}$. The recoil is much larger, up to thousands $\rm{km\,s^{-1}}$, for  MBHs with large spins in non-aligned configurations \citep{Campanelli2007,Gonzalez2007, Herrmann2007}. 

Finally, the spin of a hole might determine how much energy is extractable from the hole itself  (Blandford \& Znajek 1977; Tchekhovskoy et al. 2011; McKinney et al. 2012). The so-called ``spin paradigm" asserts that powerful relativistic jets  are produced in AGNs with fast rotating black holes \citep{Blandford1990}, implying that MBHs rotate slowly in radio-quiet quasars, which represent the majority of quasars \citep{Wilson1995}. Sikora et al. (2007) proposed a ``spin-accretion paradigm", suggesting that  the production of powerful relativistic jets is conditioned by the presence of fast rotating holes, while it also depends on the accretion rate and on the presence of disk magneto-hydrodynamical winds required to provide the initial collimation of the central Poynting flux dominated outflow, as in, e.g., the Blandford-Znajek process. Recently Sikora \& Begelman (2013) proposed that the magnetic flux threading the black hole, rather than BH spin or Eddington ratio, is the dominant factor in launching powerful jets.

As described above, MBH spins determine directly the mass-to-energy conversion efficiency of quasars. On the other hand, accretion determines the evolution of MBH spins.  A hole that is initially non-rotating  gets spun up to a maximally-rotating state ($a=1$) after increasing its mass by a factor $\sqrt{6}\simeq 2.4$.     A maximally-rotating hole is spun down by retrograde accretion to $a=0$ after growing by a factor $\sqrt{3/2}\simeq 1.22$. Different modes of MBH feeding imply different spin histories. Spin-up is a natural consequence of prolonged disk-mode accretion: any hole that increases substantially its mass by capturing material with constant angular momentum axis would ends up spinning rapidly (``coherent accretion"). Spin-down occurs when counter-rotating material is accreted, if the angular momentum of the accretion disk is strongly misaligned with respect to the direction of the MBH spin.  It has been suggested that accretion may proceed also via small (i.e., the accreted mass is a very small fraction of the MBH mass, $\sim$ 1\% or less) and short uncorrelated episodes \citep[``chaotic accretion",][]{Moderski1996,King2006}, where accretion of co-rotating (causing spin-up) and counter-rotating (causing spin-down) is equally probable.  
As the ISCO for a retrograde orbit is at larger radii than for a prograde orbit, the transfer of angular momentum is more efficient in the former case. Accretion of counter-rotating material therefore spins MBHs down more efficiently than co-rotating material spins them up. \cite{King2008} considered a MBH evolution scenario where chaotic accretion very rapidly adjusts the hole's spin parameter to average values $a \sim 0.1-0.3$  from a broad range of initial conditions, only weakly dependent on the overall angular momentum distribution of the accreting gas parcels.

MBH-MBH mergers also influence the spin evolution.  \cite{BertiVolonteri2008} consider how the dynamics of BH mergers influences the final spin. Except in the case of aligned mergers, a sequence of BH mergers can lead to large spins $ >0.9$ {\it only if} MBHs start already with large spins {\it and} they do not experience many major mergers.  Therefore, the common assumption that mergers between MBHs of similar mass always lead to large spins needs to be revised.

The focus of this paper will be on the cosmic evolution of spins of massive black holes, $M \sim 10^6-10^9\,M_\odot$ \citep{richstone1998, Ferrarese2005}, specifically on how accretion and MBH-MBH mergers determine the magnitude of spins. Very few works to-date have studied the joint MBH mass and spin coevolution \citep{Moderski1998,Volonterietal2005,2005ApJ...620...59S,LPC09,Fanidakis11,Barausse12}. In common with previous efforts we adopt a semi-analytical approach, in order to capture both the cosmic evolution of structures and the processes that occur near MBHs. This approach allows us to model accretion processes using an analytical formalism, that in principle has unlimited spatial resolution. This is particularly relevant as the physical processes that influence spin evolution occur near the MBH, and unfortunately direct cosmological simulations at sub-pc resolution are still unfeasible. The other advantage of this approach is that each assumption is clearly described mathematically, making the calculation easily reproducible, or modifiable and testable under different assumptions by scientists with different theoretical stances. Finally, one should appreciate that our formalism does not have many more ``cranks" than sub-grid prescriptions adopted in numerical simulations, while offering a clear framework that can be replicated, or modified, in a very economical way using a standard desktop by any scientist who decides so. It is important to notice that our model does reproduce a large number of observational constraints (luminosity function of AGN and mass function of MBHs, relation between MBHs and hosts, mass density in MBHs at low and high-redshift). Since we are comparing our models to a large number of observables, there is not much leverage for the model parameters or assumption to be varied.   In section 5 we discuss how the model's parameters can be changed (and which cannot be modified). Of course, since there are not many observational constraints on MBH spins (but see section 6 and 7) the possibilities to compare our models with observations are rather limited. Within the assumptions made, and the observational constraints used to anchor our calculation, the model is robust.  Being this a theoretical investigation, we present a framework that predicts a set of properties for the MBH population. In contrast with previous investigations that focused on high-redshift quasars (e.g. Volonteri et al. 2005; Shapiro 2005) our main interest here is to study the populations of low-redshift MBHs and AGN whose spins may be directly measured through X-ray spectroscopy, or indirectly estimated through their average radiative efficiency. In particular, our models aim at translating the theoretical expectations in a framework that can be directly applied to observational samples, for instance by casting our results in terms of AGN luminosity rather than MBH mass as typically done in the literature (as only a small subsample of AGN have mass measurements).

The outline of the paper is as follows. In section 2 we describe the basic infrastructure that we use to model the cosmic evolution of structures. In section 3 we summarize how we model spin evolution in MBH-MBH mergers, while in section 4 we describe how different phases of accretion, related to the cosmic evolution of galaxies and of the MBHs they host, influence MBH spins. In section 5 we consider a series of observational constraints that we adopt to anchor our model. In section 6  we discuss our results, and we present our conclusions in section 7. 

\section{The backbone: dark matter halos and galaxies}
We investigate the evolution of MBHs via cosmological realizations of the  merger hierarchy of dark matter halos from early times to the present in a $\Lambda$CDM cosmology \citep[WMAP5,][]{WMAP5}. We track the merger history of 300 parent halos with present-day masses in the range $10^{11}<M_{\rm h}<10^{15}\,\msun$ with a Monte Carlo algorithm \citep{VHM}. The mass resolution of our algorithm reaches $10^5\msun$ at $z=20$, and the most massive halos are split into up to 600,000 progenitors.  

We wish to keep our models as simple as possible, while making sure that the properties of the MBHs we study are correctly determined through the cosmic evolution of their hosts. We do not explicitly model the evolution of the baryonic component of the host galaxies through cooling, star formation and various feedbacks \citep[see][and references therein for models that treat in detail semi-analytically the baryonic component of galaxies and its link to MBH evolution]{LPC09,Fanidakis11,Barausse12}. In our models we use only one parameter to link the host halo to the central MBH, and it is the host's central velocity dispersion. We link the central stellar velocity dispersion of the host to the  asymptotic virial velocity ($V_c$) assuming a spherical, isothermal halo, so that $\sigma=V_c/\sqrt[]{2}$. We calculate the circular velocity from the mass of the host halo and its redshift.   A halo of mass $M_{\rm h}$ collapsing at redshift $z$ has a circular velocity:
\beq V_c= 142 \kmps \left[\frac{M_{\rm h}}{10^{12} \ M_{\sun} }\right]^{1/3} 
\left[\frac {\Omm}{\Ommz}\ \frac{\Delta_{\rm c}} {18\pi^2}\right]^{1/6} 
(1+z)^{1/2},  
\eeq 
where $\Delta_{\rm c}$ is the over--density at virialization relative 
to the critical density. 
For a WMAP5 cosmology we adopt here the  fitting
formula  $\Delta_{\rm c}=18\pi^2+82 d-39 d^2$ 
 \citep{Bryan1998}, 
where $d\equiv \Ommz-1$, and $\Ommz \, =\,  \Omm (1+z)^3/(\Omm (1+z)^3+\Oml+\Omk (1+z)^2) $.

At high redshift we seed dark matter halos with MBHs created by gas collapse. Specifically, we adopt here the formation model detailed in Natarajan \& Volonteri (2011) based on Toomre instabilities (Lodato \& Natarajan 2006).  The Toomre parameter is defined as $Q=\frac{c_{\rm s}\kappa}{\pi G \Sigma}$, where $\Sigma$ is the  surface mass density, $c_{\rm s}$ is the sound speed,  $\kappa=\sqrt{2}V_c/R$ is the epicyclic frequency, and $V_c$  is the circular velocity of the disk. When $Q$ approaches a critical value, $Q_c$, of order unity, the disk is subject to gravitational instabilities. If the destabilization of the system is not too violent, instabilities lead to mass infall instead of fragmentation into bound clumps and global star formation in the entire disk (Lodato \& Natarajan 2006).  This process stops when the amount of mass transported to the center is sufficient to make the disk marginally stable. The mass that has to be accumulated in the center to make the disk stable, $M_{inf}$, is obtained by requiring that $Q=Q_c$. This condition can be computed from the Toomre stability criterion and from the disk properties, determined from the dark matter halo mass, via $T_{\rm vir}\propto M_h^{2/3}$, and angular momentum, via the spin parameter, $\lambda_{\rm spin}$:
\begin{equation}
M_{inf}= f_{\rm d}M_{\rm halo}\left[1-\sqrt{\frac{8\lambda_{\rm spin}}{f_{\rm d}Q_{\rm c}}\left(\frac{j_{\rm d}}{f_{\rm d}}\right)\left(\frac{T_{\rm gas}}{T_{\rm vir}}\right)^{1/2}}\right]. 
\label{mbh}
\end{equation}
for $\lambda_{\rm spin}<\lambda_{\rm max}=f_{\rm d}Q_{\rm c}/8(f_{\rm d}/j_{\rm d}) (T_{\rm  vir}/T_{\rm gas})^{1/2}$. Here $\lambda_{\rm max}$ is the maximum
halo spin parameter for which the disk is gravitationally unstable,  $f_d=0.05$ is the gas fraction that participates in the infall, and $j_d=0.05$ is the fraction  of the halo angular momentum retained by the collapsing gas.  We further assume $T_{\rm gas}=5000$K and $Q_{\rm c}=2$ (see Volonteri, Lodato \& Natarajan 2008 for a validation of the parameter choice). Given the mass and spin parameter of a halo, the mass that accretes to the center in order to make  the disk stable, $M_{inf}$, is an upper limit to the mass that can go into MBH formation. We assume here $M_{seed}=M_{inf}$. Please refer to Natarajan \& Volonteri (2011) and references therein for details on the MBH formation process. 

Our model for MBH growth requires only to know whether a galaxy is gas-rich (we typically refer to gas-rich galaxies as ``disks" in this paper) or gas-poor (``spheroid"). Since forming galaxy disks even in high-resolution cosmological simulations is extremely challenging  \citep{Governato2007}, we keep our model for galaxy morphology as simple as possible. Morphology is related  to the merger history, using a three-parameter model, where spheroid formation depends on both halo mass ratio and the absolute halo mass, and a spheroid can re-acquire a disk through cold flows and mergers with gas-rich galaxies. \cite{Koda2007} show that the fraction of disk- vs spheroid-dominated galaxies is well explained if the only merger events that lead to spheroid formation have mass ratio $>$0.3 and  virial velocity   $>55 \kms$; also, the merger timescale must be inferior to the time between when the merger starts and today, $z=0$. We assume that spheroids form after a merger that meets these requirement.   We additionally allow a disk to reform after 5 Gyrs in galaxies with virial velocity $<300 \kms$ where no major mergers occurred to include the effect of cold flows.  In section 5 we discuss the sense of this approach. 

\section{Spin evolution due to MBH-MBH mergers}
We  assume that, when two galaxies hosting MBHs merge, the MBHs themselves merge within the merger timescale of the host halos \citep[and references therein]{Sesana2007,Dotti2007}. We adopt the relations suggested by \cite{Boylan2008} for the galaxy merger timescale.  We model MBH spin changes due to mergers adopting an analytical scheme similar to that described in Berti \& Volonteri (2008), based on simulations of black hole mergers in full general relativity \citep{Rezzolla2008, Campanelli2009}. \cite{Kesden2010} has validated the consistency of different fitting formulae for calculating the spin of MBH remnants, and we refer the reader to \cite{Lousto2010} for the most comprehensive fitting formulae. Due to computational constraints, we adopt here the fitting formulae of Rezzolla et al. (2008) for their easy implementation:

\begin{eqnarray}
\label{eq:general}
&\hskip -0.5cm \quad \vert \boldsymbol{a}_{\rm fin}\vert=
\frac{1}{(1+q)^2}\Big[ \vta{1}^2 + \vta{2}^2 q^4+
 2 {\vert \boldsymbol{a}_2\vert}{\vert 
\boldsymbol{a}_1\vert} q^2 \cos \alpha +
\nonumber\\ 
& \hskip 1.cm 
2\left(
     {\vert \boldsymbol{a}_1\vert}\cos \beta +
     {\vert \boldsymbol{a}_2\vert} q^2  \cos \gamma
\right) {\vert \boldsymbol{{\ell}} \vert}{q}
+ \vert \boldsymbol{{\ell}}\vert^2 q^2
\Big]^{1/2}\,,
\end{eqnarray}
where $q\equiv
M_2/M_1\leqslant 1$ is the mass ratio between the two MBHs; the three (cosine) angles $\alpha, \beta$ and $\gamma$ are 
defined by
\begin{equation}
\label{cosines}
\cos \alpha \equiv
{\boldsymbol{\hat{a}}_1\cdot\boldsymbol{\hat{a}}_2}
\,,
\hskip 0.3cm
\cos \beta \equiv
 \boldsymbol{\hat a}_1\cdot\boldsymbol{\hat{{\ell}}}\,,
\hskip 0.3cm
\cos \gamma \equiv
\boldsymbol{\hat{a}}_2\cdot\boldsymbol{\hat{{\ell}}}\,.
\end{equation}
where $\vtl$ is the magnitude of the  orbital angular momentum, and

\begin{eqnarray}
\label{eq:L}
&&
\hskip -0.5cm
\vtl
 = 
\frac{s_4}{(1+q^2)^2} \left(\vta{1}^2 + \vta{2}^2 q^4 
	+ 2 \vta{1} \vta{2} q^2 \cos\alpha\right) + 
\nonumber \\
&& \hskip 0.5cm
\left(\frac{s_5 \nu + t_0 + 2}{1+q^2}\right)
	\left(\vta{1}\cos\beta + \vta{2} q^2 \cos\gamma\right) +
\nonumber \\
&& \hskip 0.5cm
	2 \sqrt{3}+ t_2 \nu + t_3 \nu^2 \,,
\end{eqnarray}

where $\nu$ is the symmetric mass ratio $\nu \equiv
M_1M_2/(M_1+M_2)^2$, and  the coefficients take the values $s_4 =
-0.129 \pm 0.012$, $s_5 = -0.384 \pm 0.261$, $t_0 = -2.686 \pm 0.065$,
$t_2 = -3.454 \pm 0.132$, $t_3 = 2.353 \pm 0.548$.

We model the spin--orbit configuration differently depending on the properties of the host galaxies. During a gas-rich merger, large amounts of gas are driven toward the centers of the two interacting galaxies \citep{Downes1998} and form a dense circumnuclear disk in which the MBHs settle. In this phase, the MBHs accrete in a coherent manner at a rate sufficient to {\it align} their spins, initially oriented at random, to the angular momentum of the nuclear disk 
\citep{Liu2004,Bogdanovic2007,Dotti2010}: in response to the Bardeen--Petterson \citep{1975ApJ...195L..65B} warping of the small--scale accretion disks grown around each MBH, total  angular momentum conservation imposes fast ($\lsim 1$ Myr) alignment of the BH spins with the angular momentum of their orbit and so of the large--scale circumnuclear disk \citep{1996MNRAS.282..291S,NatarajanPringle1998,1999MNRAS.309..961N,2000MNRAS.315..570N,VSL07,2009MNRAS.400..383M,2010MNRAS.405.1212L,2013MNRAS.428.1986U} unless one considers a vertical viscosity equal to the radial one as in early works \citep{1983MNRAS.202.1181P,1985MNRAS.213..435K}, or discs with low viscosity, such as proto-planetary discs or thick discs \citep{2002MNRAS.337..706L}. See the Appendix  for additional information.

Thereafter, accretion remains prograde until coalescence, with no major changes in the MBH spin orientation.  Under these circumstances,  $\cos\alpha=\cos\beta=\cos \gamma=1$, and the MBH remnant retains the spin direction of the parent MBH spins, both oriented parallel to the angular momentum of their orbit: the post-coalescence MBH may thus acquire a large spin $> 0.7-0.9$ (Berti \& Volonteri 2008), sum of the internal and orbital spins. 

In the case of gas-poor mergers, instead, i.e. when MBHs do not accrete during mergers, evolving solely via stellar dynamical processes, there is no reason to expect any symmetry or alignment \citep{Bogdanovic2007}, so isotropy should be a good assumption in the absence of accretion disks or gas, so that $\cos\alpha$, $\cos\beta$, and $\cos \gamma$ are isotropically distributed.  \cite{BertiVolonteri2008} show that for isotropic configurations  mergers tend to ``spin-down'' a fast-spinning hole \citep[see also][]{HB2003}. For intermediate-large mass ratios (mass ratio $q=M_{\rm BH,2}/M_{\rm BH,2}\leqslant 1$ between 0.1 and 1) mergers tend to produce MBHs with average spins very close to the value $\simeq 0.7$ resulting from equal-mass, non-spinning mergers.  For smaller mass ratios the larger MBH dominates the dynamics, and the final spin can be substantially larger or smaller than this value.  

\section{Spin evolution due to accretion}
We discuss here the feeding of MBHs in the quasar phase and its aftermath and in the more quiescent Seyfert galaxies. Simulations of galaxy mergers and MBH activity \citep{DiMatteo2005,Hopkins2006} show that  for every accretion episode triggered by a galaxy merger  a three phase picture can be drawn. At the beginning the MBH has an ``healthy diet", with $f_{\rm Edd}\equiv L/L_{\rm Edd} \leqslant 1$. When the MBH mass reaches the ``M-$\sigma$" relation, the MBHs feedback can be sufficient to unbind and ``blowout'' the gas feeding it \citep{HopkinsHernquist2006}, causing a final ``starvation'', when $f_{\rm Edd}$ rapidly decreases, until no more gas is available to feed the MBH. We argue that during the healthy diet and blowout  phases MBHs gain a high spin while accreting efficiently and coherently \citep{Dotti2010},  building the population of high-{\it z} quasars. Coherent accretion ensues because MBHs in merger remnants are expected to be surrounded by dense circum-nuclear disks \citep{SandersMirabel1996}. Maio et al. (2012) study the evolution of the angular momentum of material feeding MBHs embedded in circumnuclear disks, and they find that  coherence of the accretion flow near each MBH reflects the large-scale coherence of the disk's rotation. 

After the ``blowout" phase, starving MBHs are no longer surrounded by a thick gas disk, that determines the angular momentum of the material ending up in the accretion disk.  During this last phase we do not expect accretion to necessarily proceed coherently any longer.  
During the starvation phase the accretion rate decreases rapidly.   The depletion of gas in galaxies and the decrease in the galaxy interaction rate at late cosmic times causes therefore a widespread ``famine" in low-{\it z}  ellipticals (i.e., the merger remnants), where AGN with low accretion rates dominate  \citep[``radio" mode, see, e.g.,][] {Croton2006,Churazov2005}. 


We investigate the evolution of MBH spins during the quasar phase expanding previous work \citep{Volonterietal2005, VSL07} to more realistic models. We model the joint mass and spin evolution by coupling the results on the mass accretion rate as a function of time in simulations \citep{hopkins2005a, VSH2006}, with the spin evolution due to disk accretion \citep{VSL07}, thus solving a system of two coupled differential equations ($f_{\rm Edd}$ as a function of $M$ and time; spin $a$  as a function of $f_{\rm Edd}$, $M$. The framework has been derived in Volonteri et al. 2005, Volonteri et al. 2006, Hopkins \&  Hernquist 2006, and Volonteri et al. 2007). We remind here the relevant information.

\subsection{Quasar phase}
After a halo merger with mass ratio larger that 3:10, in which at least one of the two is a disk galaxy (hence, with conspicuous cold gas content) we assume that a merger-driven accretion episode is triggered\footnote{Please note that this assumption is at variance with previous models of MBH cosmic evolution within our framework \citep[e.g.,][]{VHM, VLN2008,VN09}, where the threshold was set to a lower value of 1:10. The reason for increasing the threshold for QSO phase to occur is the addition of  avenues for MBH growth other than merger-driven accretion and MBH-MBH mergers.}. After a dynamical timescale is elapsed  \citep[roughly, after the first pericentric passage, cf.][]{svanwas2012}  accretion starts. If at that point the MBH mass lies below the M-$\sigma$ relation, accretion occurs at the Eddington rate ($f_{\rm Edd}=1$). Additionally, during this early phase the MBH is nested into a nuclear disk that feeds the MBH coherently. The following scheme is applied to the joint evolution of mass, spin and radiative efficiency in this phase. Let us define $M$ and $a$ as the black hole mass and spin parameter at the beginning of the timestep, and $\mu$ as the cosine of the angle between the MBH spin and the inner accretion disk angular momentum.  Irrespective of the infalling material's original angular momentum vector, Lense-Thirring precession imposes axisymmetry close in, with the gas accreting on either prograde ($\mu=1$)  or retrograde equatorial orbits ($\mu=-1$).  In natural units, where $c=G=1$:
\beq
r_{\rm ISCO} = 3 + Z_2 \mp\sqrt{(3 - Z_1)(3 + Z_1 + 2 Z_2)},
\label{isco}
\eeq
is the radius of the ISCO, where $Z_1$ and $Z_2$ are functions of $a$ only (Bardeen, Press, \& Teukolsky 1972),
\begin{eqnarray}
Z_1 &\equiv & 1 + (1 - a^2)^{1/3} [(1 + a)^{1/3} + (1 - a)^{1/3}],\\
Z_2 &\equiv & [3 a^2 + Z_1^2]^{1/2},
\end{eqnarray}
and the upper (lower) sign refers to prograde (retrograde) orbits.  We calculate the accretion efficiency as:
\begin{eqnarray}
\epsilon&=&1-E_{\rm ISCO},\\
E_{\rm ISCO}&=&\left(1-{\frac{2}{3 r_{\rm ISCO}}}\right)^{1/2},
\end{eqnarray}
which is is also a plausible assumption for the radiative efficiency ($\epsilon$, mass-to-energy conversion) for thin-disk accretion occurring at large fractions of the Eddington rate (see below for the case of radiatively inefficient flows). We calculate self-consistently  the radiative efficiency from the MBH spin and the location of the ISCO (i.e., taking into consideration the direction of the relative angular momentum of spin and disk, co- or counter-rotating).

Assuming that during a timestep $\Delta t \sim 10^5-10^6$ yr the radiative efficiency and Eddington rate remain constant, ($\epsilon$=const and $f_{\rm Edd}$=const,) from the derivation shown in the  Appendix, one obtains that the MBH mass grows as:
\beq
M(t+\Delta t)=M(t)  \exp\left({f_{\rm Edd}\frac{\Delta t}{t_{\rm
      Edd}}\frac{1-\epsilon(t)}{\epsilon(t)}}\right)
\eeq
where $t_{\rm Edd}=\frac{\sigma_T \,c}{4\pi \,G\,m_p}= 0.45$ Gyr  and $f_{\rm Edd}$ represents the Eddington fraction. We update the magnitude of the MBH spin through:
\begin{eqnarray}
a(t+\Delta t)= { \frac{r_{\rm ISCO}(t)^{1/2}}{3}} {\frac{M(t)}{M(t+\Delta t)}} &&\\
\left[4-\left({{\frac{3M(t)^2} {M(t+\Delta t)^2}}} r_{\rm ISCO}(t) -2\right)^{1/2} \right] \nonumber\\
{\rm for}~{M(t+\Delta t)\over M(t)}\le  r^{1/2}_{\rm ISCO}(t), \nonumber\\
a(t+\Delta t)= 0.998~~{\rm for}~{M(t+\Delta t)\over M(t)}\ge r^{1/2}_{\rm ISCO}(t)&&
\label{spinmag}
\end{eqnarray} 
(Bardeen 1970)\footnote{We limit the MBH spin to $a=0.998$ following the calculation of Thorne (1974) that showed that the radiation emitted by the disk and swallowed by the hole produces a counteracting torque, which prevents spin up beyond this value.  We note that magnetic fields connecting material in the disk and the plunging region may further reduce the equilibrium spin by transporting angular momentum outward in non-geometrically thin disks. Fully relativistic magnetohydrodynamic simulations for a series of thick accretion disk models show that spin equilibrium is reached at $a\approx 0.93$ \citep{Gammie2004}, while in slim disks accretion can  (for low viscosities) increase the spin up to $a=0.9994$  \citep{Sadowski2011}}.  After updating the spin magnitude, we also update the mass-to-energy conversion efficiency by determining the new ISCO corresponding to $a(t+\Delta t)$, and therefore  $\epsilon(t+\Delta t)$ to be used at the successive timestep iteratively. We here assume fast alignment between accretion disk and spin  \citep[see][]{NatarajanPringle1998,Volonterietal2005,VSL07, Perego2009}, as the alignment timescale is $\simeq$ Myr, so that accretion is prograde during the quasar phase. 

%

\subsection{Decline phase}
When a MBH reaches a mass close to the value corresponding to the M-$\sigma$ correlation ($M_{BH,\sigma}$) for its host, we assume, following \cite{Hopkins2006,HopkinsHernquist2006} that a self-regulation ensues and the MBH feedback unbinds the gas closest to the MBH, thus reducing its feeding. For simplicity we  further assume that the black hole--$\sigma$ ($M$--$\sigma$) scaling is:
\beq
M \, =\, 10^8  \left(\frac{\sigma}{200 \kmps} \right)^4 \, \msun
\eeq
 \citep{Tremaine2002}. 
To match the luminosity function of quasars, we start the decline phase when the MBH mass is $0.25\times M_{BH,\sigma}$. The MBH continues accreting during the decline of accretion and it typically reaches a value closer to the $M_{BH,\sigma}$ by the end of the accretion episode.  Note that  our model  does not necessarily imply that the $M_{BH,\sigma}$ relation is a tight correlation. We assume  only that  the feedback during a high-accretion rate quasar phase establishes  at that time for that object an $M$--$\sigma$ relation (cf. Silk \& Rees 1998; Fabian 1999). Additional processes, such as MBH-MBH mergers (sec. 3), accretion during the ``decline phase" (sec. 4.2), accretion of recycled gas (sec. 4.3), accretion of gas stolen from molecular clouds (sec. 4.4) do not have any limit imposed by the $M_{BH,\sigma}$ relation, and in fact they produce scatter, by pushing the MBHs above or below the relationship, depending also on the galaxy history (see Volonteri \& Ciotti 2012). 

We model the decrease of the accretion rate through the analytical formula   from \cite{HopkinsHernquist2006}:
\beq
f_{\rm Edd}(t)=\left(\frac{t+t_{f_{\rm Edd}}}{t_{f_{\rm Edd}}}\right)^{-\eta_L},
\label{decay}
\eeq
where $\eta_L \simeq2$ and $t_{f_{\rm Edd}}\simeq4.1\times 10^6(M_{BH,\sigma}/10^8)\msun$ yr, and $t=0$ (where $f_{\rm Edd}=1$) represents the time when the MBH reaches the threshold ($0.25\times M_{BH,\sigma}$).  Since $\dot{M}_{\rm in}(t)=f_{\rm Edd}(t)\,M(t)/t_{\rm Edd}$ ,  depends on time, the accretion rate must be integrated self-consistently, and the mass now grows with time as:
\begin{eqnarray}\nonumber
M(t) & =& M(0)  \exp \left( \frac{1}{\eta_L-1}\left[ \frac{1}{t^{1-\eta_L}_{f_{\rm Edd}}} -\frac{1}{(t+t_{f_{\rm Edd}})^{1-\eta_L}}\right]  \right.\\
&& \left. \frac{t^{\eta_L}_{f_{\rm Edd}}}{t_{\rm Edd}}\frac{1-\epsilon(t)}{\epsilon(t)}  \right).
\label{decaym}
\end{eqnarray}

We use again Eq.~\ref{isco}--\ref{spinmag} to model spin evolution, however, we explore two possible scenarios. In our reference case we assume that the ``outflow" causes some stirring of the angular momentum of the gas within the central region. We therefore explore a ``chaotic" case where in the decline phase we  pick a new random $\mu=1$ or $\mu=-1$ at each timestep ($\simeq 10^5-10^6$ years) to mimic the lack of coherence in the accretion flow after MBH feedback has blown away the surrounding gas. In a second model we assume instead that  the ``blow-out" of gas does not  affect strongly the angular momentum of the material near the MBH, and persist with keeping $\mu=1$.   
Given that the timescale for decline is longer for larger MBHs (cf. Eq.~\ref{decay}), the larger the MBH the longer the phase at relatively high accretion rates, and the faster the decrease of MBH spin, as more mass is accreted in a non-coherent fashion.  

When the accretion rates become very sub-Eddington, we assume that the accretion flow becomes optically thin and geometrically thick.  In this state the radiative power is strongly suppressed \citep[e.g.,][]{Narayan1994,Abramowicz1995}, so that the radiative efficiency differs from 
the  mass-to-energy conversion efficiency, $\epsilon$, that depends on the location of the ISCO only. Indeed, the radiative efficiency becomes very model dependent and uncertain. In order to estimate the effect of radiatively inefficient accretion on the MBH population we adopt here a specific functional form for the radiative efficiency.   Following \cite{Merloni08} we write the radiative efficiency, $\epsilon_{\rm rad}$, as a combination of the mass-to-energy conversion, $\epsilon$, and of a term  that depends on the properties of the accretion flow itself.  \cite{Merloni08} suggest that the transition in the disk properties occurs\footnote{Please note that  \cite{Merloni08} use a different notation and terminology. Their $\lambda$ is our $f_{\rm Edd}\equiv L/L_{\rm Edd}\equiv \epsilon_{\rm rad} \dot{M} c^2/L_{\rm Edd}$, and their $\dot{m}=\epsilon\,f_{\rm Edd}/\epsilon_{\rm rad}$. As long as the accretion flow is optically thick and geometrically thin, i.e., before the transition to very sub-Eddington flows, $\dot{m}=f_{\rm Edd}$.} at $f_{\rm Edd}<f_{\rm Edd,cr}=3\times 10^{-2}$, and that $\epsilon_{\rm rad}=\epsilon\,(f_{\rm Edd}/f_{\rm Edd,cr})$.  This specific choice allows us to estimate  qualitatively the impact of radiatively inefficient sources to the AGN populations, and on the inferences that one can (or not) make from observables.  

\subsection{Quiescent elliptical phase}
After the formation of an elliptical galaxy,  the feeding of the MBH can be sustained by the recycled gas (primarily from red giant winds and planetary nebulae) of the evolving stellar population \citep{Ciotti1997,Ciotti2001,Ciotti2007,Ciotti2010}. As shown by  \cite{Ciotti2011} the behaviour of the accretion rate is similar to what we describe above: at early times the evolution is characterized by major, albeit intermittent, accretion episodes, while at low redshift accretion is smooth and characterized by $f_{\rm Edd}\ll1$. We have implemented this channel of MBH feeding only for quiescent ellipticals, that is only after a spheroid is formed and had time to relax.   When the MBH mass is $\sim 10^{-3}$ of the spheroid mass \citep{Magorrian1998,MarconiHunt2003,Haring2004}, and the spheroid is modeled as a Hernquist profile, the geometrical model by \cite{Volonteri2011} implies that the quiescent level of accretion onto a central MBH due to recycled gas is $f_{\rm Edd}\simeq 10^{-5}$. We  assume here that this mode of accretion has a constant $f_{\rm Edd}=10^{-5}$, and we model spin evolution assuming that that the accreted material is isotropically distributed (i.e.,  we pick a random $\mu=1$ or $\mu=-1$ at each timestep, and use Eq.~\ref{spinmag} to evolve the spin magnitude). This phase has little effect on the magnitude of MBH spins, due to the low accretion rates and the modest implied MBH growth (cf. Eq.~\ref{spinmag}).

\subsection{Molecular cloud accretion in disk galaxies}
Several observations suggest that single accretion events last  $\simeq 10^5$ years in Seyfert galaxies, while the total activity lifetime (based on the fraction of disk galaxies that are Seyfert) is $10^8-10^9$ years \citep[e.g.,][]{Kharb2006,Ho1997}. This suggests that accretion events are very small and very {`}compact'. Smaller MBHs, powering low luminosity AGN, likely grow by accreting smaller packets of material, such as molecular clouds \citep{HopkinsHernquist2006}. Compact self-gravitating cores of molecular clouds (MC) can occasionally reach subparsec regions.  In gas-rich, star-forming disk galaxies the MBH is  likely to be fed by short, recurrent, uncorrelated accretion episodes.  The spin evolution of a MBH hosted by a quiescent disk galaxy would then resemble the ``chaotic accretion" scenario. This argument was discussed only qualitatively by  \cite{VSL07}. We now wish to provide quantitative statistical predictions for the distribution of MBH spins in different hosts. We follow here \cite{Sanders1981} and \cite{HopkinsHernquist2006} to determine the event rate, and Volonteri et al. 2007 to couple accretion episodes to spin evolution. 
 
In a disk galaxy, at each timestep, $\Delta t$, we determine the probability of a MC accretion event as:  
\beq
{\cal P}=\frac{\Delta t}{t_{MC}}\simeq \frac{10^{-3} \sigma {\Delta t}}{R_{\rm cl}}
\label{prob}
\eeq
where $R_{\rm cl}\simeq 10$ pc  (Hopkins \& Hernquist 2006).  As in \cite {VSL07} we further assume a lognormal distribution (peaked at $\log(M_{\rm MC}/$\msun$)=4$, with a dispersion of 0.75) for the mass function of MC close to galaxy centers \citep[based on the Milky Way case, e.g.,][]{Perets2007}. 

We model accretion of MCs through a description inspired by   Bottema \& Sanders (1986) and Wardle \& Yusef-Zadeh (2008). We assume that the MBH captures only material passing within the Bondi radius, $R_B$, and we also assume that specific angular momentum is conserved, so that the outer edge of the disk that forms around the MBH corresponds to the material originally at the Bondi radius:
\beq
R_d=2 \lambda^2 R_B=8.9\, {\rm pc}\,\lambda^2 \frac{M}{10^7\msun}\left(\frac{\sigma}{100 \kms}\right)^{-2}, 
\eeq
where $\lambda$ is the fraction of angular momentum retained by gas during circularization. The maximum captured mass will be contained in a cylinder with radius $R_B$ and length $2\times R_{\rm cl}$, the MC diameter. If $\kappa$ is the ratio of the mass going into the disk with respect to  the whole mass in the cylinder, then:
\begin{eqnarray}\nonumber
M_{\rm d,max}&=& \kappa \frac{R_B}{R_{\rm cl}}M_{\rm MC}= 4.7\times 10^4  \kappa  \left(\frac{M}{10^7\msun}\right)^2\\
&& \left(\frac{\sigma}{100 \kms}\right)^{-4} \msun. 
\label{MdiscMC}
\end{eqnarray}
The inflow time will be of order of the viscous timescale for the disk, 
\begin{eqnarray}\nonumber
t_{\rm visc}&=&\left(\frac{R^{3}_d}{\alpha_v^2 GM_d}\right)^{1/2}=3.4 \times 10^5 \frac{\lambda^3}{\alpha_v}\frac{M}{10^7\msun}\\
&& \left(\frac{\sigma}{100 \kms}\right)^{-3}  {\rm yr},
\end{eqnarray}
so that for the whole disk to be consumed we can calculate an upper limit to the mean accretion rate and luminosity:
\begin{eqnarray}\nonumber
\dot{M}_{\rm max}&=&\frac{M_{\rm d,max}}{t_{\rm visc}}=0.13 \frac{\alpha_v \kappa}{\lambda^3}\frac{M}{10^7\msun}\\
&& \left(\frac{\sigma}{100 \kms}\right)^{-1} N_{23} \,\msun {\rm yr}^{-1} , 
\end{eqnarray}
where $N_{23}$ is the column density in the MC in units of $10^{23}$ cm$^{-2}$, and $\alpha_v=0.1$ is the viscosity parameter.  Following Wardle \& Yusef-Zadeh (2011) we  set $\lambda=0.3$ and $\kappa=1$. If $\dot{M}_{\rm max}$ is less than the Eddington rate (assuming a radiative efficiency of 10\%) we let the MBH accrete the whole $M_{\rm d,max}$ over a time $t_{\rm visc}$, otherwise we treat accretion similarly to the ``decline" phase of quasars (Eq.~\ref{decay} and \ref{decaym}), as feedback from the high luminosity produced by accreting the cloud will limit the amount of material the MBH can effectively swallow.

\begin{figure}
\includegraphics[width= \columnwidth]{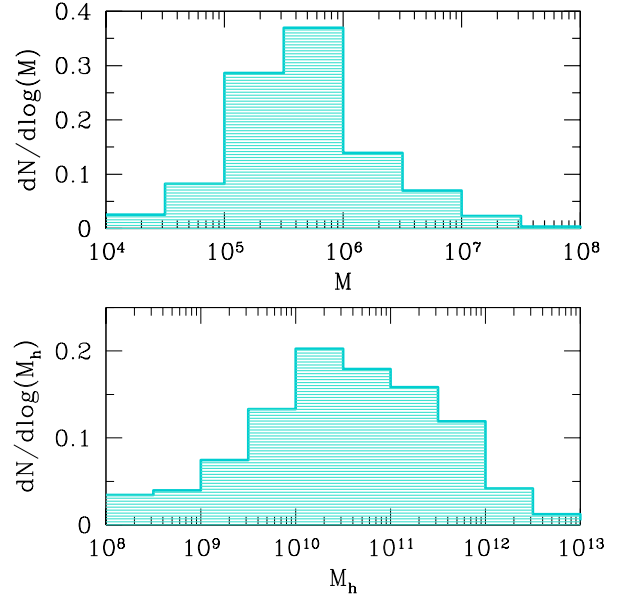}
\caption{Properties of all MBHs fed  by MCs at all redshifts: normalized distribution of the masses of MBHs  (top) and their host halos (bottom).  The probability of MC accretion increases with galaxy mass, on the other hand most massive galaxies are spheroids, and therefore they do not have a population of MCs available.}
\label{MC}
\end{figure}

From Eq.~\ref{MdiscMC}, it is evident that  the mass accreted in one of these episodes is typically much less than the mass of the MBH (typically between $10^{-2}$ and $10^{-5}$ of the MBH mass), we therefore assume that no alignment between accretion disk and MBH spin can occur, and  that retro- and prograde accretion is equally probable, i.e., we assign $\mu=1$ or $\mu=-1$ with equal probability at each event, and keep $\mu$ constant over the accretion phase, using Eq.~\ref{spinmag} to evolve the spin magnitude. As shown by Volonteri et al. (2007)  these assumptions result is a spin down in a random walk fashion that depends on the mass of the MBH and on the number of events.  We note that this is an extreme condition of randomness in the orbits and distribution of MCs, as any common sense of rotation caused by the presence of disk-like structures in the host would decrease the degree of anisotropy (Dotti et al. 2012).

Eq.~\ref{prob}, instead, shows that accretion of MCs is more probable in large galaxies, since the accretion probability is directly proportional to the velocity dispersion of the galaxy. Large galaxies, however, are more likely to be gas-poor spheroids. Large galaxies therefore have a lower probability of being gas-rich and host a population of MCs, while at the same time their MBHs have a higher probability of capturing MCs, if clouds are present. Thus, this type of accretion events occur typically in galaxies hosted in halos with mass $\sim 10^{11}-10^{12}\, \msun$ and fuel MBHs with mass $\sim 10^{5}-10^{7}\, \msun$. In Fig.~\ref{MC} we show the distribution of the masses of MBHs and their host halos where accretion of MCs takes place according to our scheme.

\section{Anchoring the model}
Here we present the constraints that we use to ``anchor" our model to observed properties of galaxies and AGN. After validating our scheme for accretion and host evolution, we will discuss what its implications are for the, still unknown, distribution of MBH spins. 

First, Fig.~\ref{morph} we compare our model to constraints at $z=0$. In the bottom panel of Fig.~\ref{morph} we show the morphological fraction as a function of galaxy stellar mass. We scale from halo mass to stellar mass through the data described in Fig.~1 of \cite{Hopkins2010}, assuming that a fixed fraction 10-50\% of the baryons is in stars (open red points and filled orange ones respectively).  While agreement is far from perfect, our simple approach reproduces the correct trend. We recall here that we do not model the whole evolution of gas and stars, nor  the disk formation \citep[cf.][for a comprehensive model]{Barausse12}, but  to derive all properties our scheme uses only a single quantity, the halo mass.

In the top panel of Fig.~\ref{morph} we reproduce the relationship between MBH masses and circular velocity \citep{VNG2011}. Circles are model MBHs at $z=0$, while errorbars show the data (Kormendy \& Bender 2011; Volonteri et al. 2011b).  We note that our model fails to reproduce some of the massive MBHs at $z=0$ in spheroids. The reason for the suppressed growth is that no MC accretion can happen in our scheme in gas-poor galaxies, and we have not implemented a time-dependent accretion of recycled gas. As noted by  \cite{Ciotti2007} the behavior of a MBH fueled by stellar mass loss is self-regulated between ``on" and ``off" phases. Our model includes only the ``off" (quiescent) phase and therefore underestimates the growth due to this fueling channel. 

Our second anchor is the luminosity function of AGN in the redshift range $0.5\leqslant z \leqslant 3$. This is a strong constraint for our accretion scheme, and it is shown in Fig.~\ref{LF}. Our scheme produces an AGN population in good agreement with the observations at $z=0.5$, $z=2$ and $z=3$, while we slightly underproduce AGN at $z=1$. The lack of high luminosity quasars at $z=2$ and $z=3$ is due to our merger trees not including halos massive enough to host MBHs with masses above a few $10^8$ \msun at those redshifts. Overall, however, we obtain the correct trends. This figure shows that there is little difference in the two models we explore on the effect of feedback over the angular momentum of the nuclear gas (``chaotic" or ``coherent" decline phases). The difference between chaotic and coherent decline (the ``quasar" phase is coherent in all cases), reflects only on spin and as a consequence on radiative efficiency, not on the Eddington ratio.  In the following we will distinguish the two models only when the results are significantly different, otherwise we show only the reference case (``chaotic" decline phase).

The main parameters influencing the performance of the model against the constraints are the mass ratio above which a mergers can trigger quasar activity and the fraction of  $M_{BH,\sigma}$ when the decline phase starts. These two parameters are weakly degenerate. The former parameter is set to $>3:10$ in the present model to match as well as possible the luminosity function at the bright end without overestimating MBH masses at a given circular velocity at $z=0$, as  happens instead by choosing a lower threshold. A much lower threshold would also be in disagreement with simulations of galaxy mergers that study merger-drive AGN activity, and show that  with a mass ratio of 1:6 high level of AGN activity does not occur (Van Wassenhove et al. in preparation). We tested a case where we instead increased the threshold to 1:2, and in this case the bright end of the luminosity function would disappear  at $z>2$ (while little or no change occurs at the faint end). For the latter parameter, we adopted a value of $0.25\times M_{BH,\sigma}$.  We  tested a case that brought the MBHs exactly  on $M_{BH,\sigma}$, but this leads to largely overestimating the MBH masses at $z=0$ at a given $V_c$. We also tested a case with $0.5\times M_{BH,\sigma}$, and in that case we still overestimated MBH masses at $z=0$ at a given $V_c$ (the overestimate is a factor of 2 overall in this case, over the best fit relationship). Decreasing the parameter value to $0.125\times M_{BH,\sigma}$ instead  underestimates  MBH masses at $z=0$ at a given $V_c$, by a factor 2.25 overall, over the best fit relationship. Finally, we tested a case where we decreased the mass ratio threshold for merger-driven AGN activity to 1:10, and at the same time we decreased the mass limit to $0.125\times M_{BH,\sigma}$ to compensate. In this case the luminosity function is similar to the case with 1:10 and $0.25\times M_{BH,\sigma}$, but the relationship between MBH mass and $V_c$ is tilted, having a shallower slope that underestimates the real MBH masses at the high $V_c$ end. If we were to choose 1:10 and $0.25\times M_{BH,\sigma}$, then the relationship between MBH mass and $V_c$ is overall overestimated by more than a factor of 2.  In summary, we have run several tests to limit the space of free parameters, and to disentangle weakly degenerate ones until we found the set that best matches the set of observational constraints.

Accretion of molecular clouds  affects the faint end of the luminosity function. One could in principle boost the probability of MC accretion by assuming more compact clouds, however the mass gained through this process is constrained by the faint end of the luminosity function of AGN (Fig.~\ref{LF}), and only small variations can be tolerated by our model, as the current implementation gives a good match with observations. We note that MC accretion accounts for almost all sources up to $L\simeq10^{12}\,$L$_\odot$ at $z=0.5-1$ and $L\simeq10^{10}\,$L$_\odot$ at $z=2-3$.}

Finally, we generate ad hoc merger trees of $M_h=2\times 10^{13}~\msun$ halos at $z=5-6$ to check that our model reproduces the existence of powerful quasars at $z\sim6$, and that the mass density we obtain at $z>5$, of $6\times10^3$ \msun/Mpc$^3$, does not overproduce the X-ray background (upper limit of $10^4$ \msun/Mpc$^3$, Salvaterra et al. 2012). Fig.~\ref{MF6} compares the theoretical mass function of MBHs that power quasars with bolometric luminosity larger than $10^{45}$ erg/s at $z=6$ to the empirical mass function derived by \cite{Willott2010} from a sample of $z=6$ quasars in the Canada--France High-z Quasar Survey. 

As discussed above, our model, while far from being able to explain every single detail of the MBH population and its growth, qualitatively grasps most of the global behavior. We therefore consider our attempts to model the spin evolution also of qualitative nature. Regardless of the simplified nature of our models, we can learn how different  patterns influence the evolution of MBH spins.

\begin{figure}
\includegraphics[width= \columnwidth]{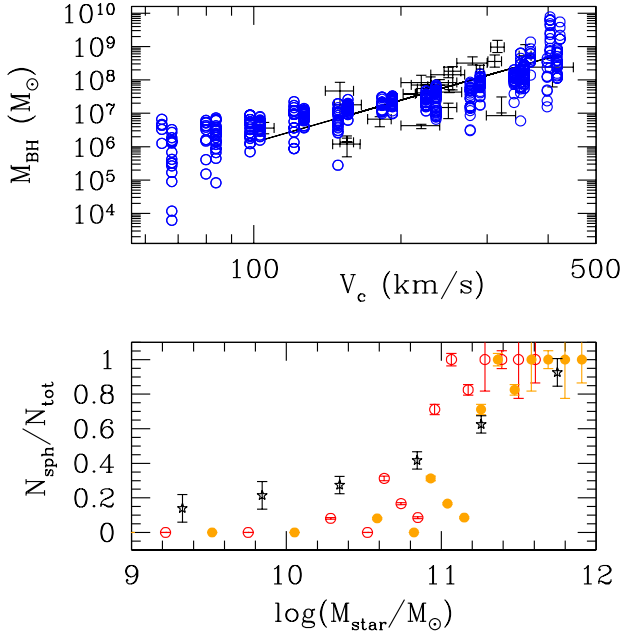}
\caption{Top panel: relationship between MBH masses and circular velocity. Circles are model MBHs at $z=0$, errorbars show the datapoints collected in Kormendy \& Bender (2011) and the best fit derived in \cite{VNG2011}. Bottom panel: fraction of spheroids as a function of stellar mass. We scale from halo mass to stellar mass through the data described in Fig.~1 of Hopkinset al. 2010, assuming that a fixed fraction 10-50\% of the baryons is in stars (open red points and filled orange ones respectively). We compare the fraction of spheroids to Conselice et al. (2006).}
\label{morph}
\end{figure}

\begin{figure}
\includegraphics[width= \columnwidth]{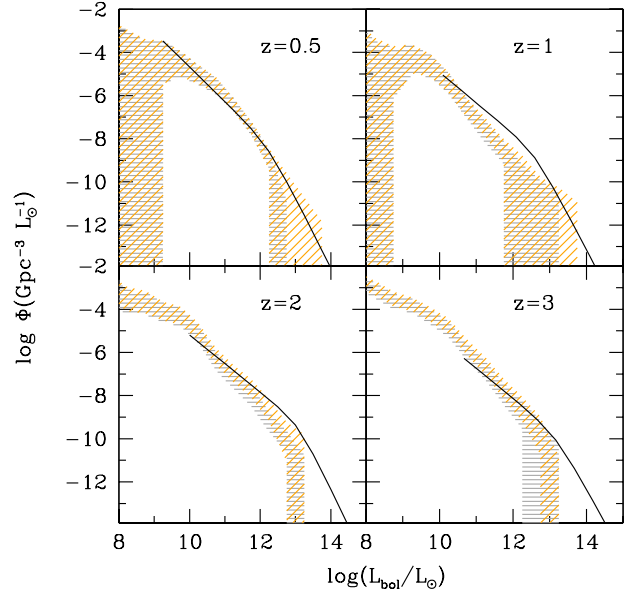}
\caption{Luminosity function at different redshifts. We show here minimum and maximum values, considering both 1-$\sigma$ statistical uncertainties, using Poissonian statistics, and fraction of absorbed AGN (La Franca et al. 2005). Orange (45$^\circ$ hatching): coherent accretion during the decline of the quasar phase. Gray (horizontal hatching):  chaotic accretion during the decline of the quasar phase.}
\label{LF}
\end{figure}

\begin{figure}
\includegraphics[width= \columnwidth]{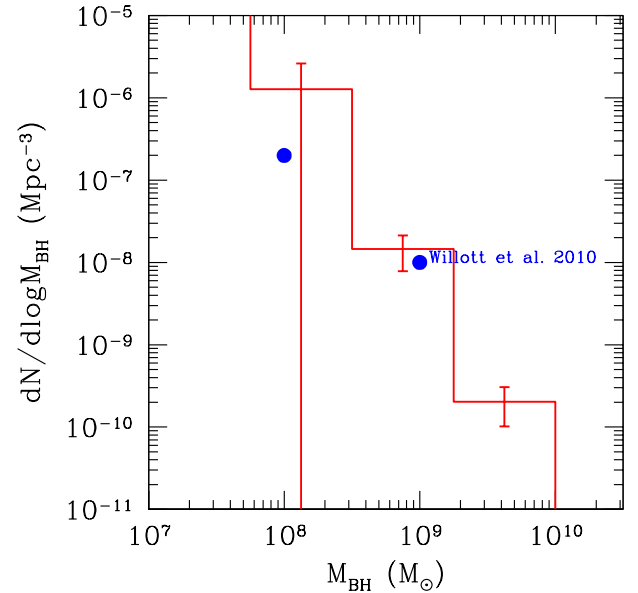}
\caption{Red histogram: theoretical mass function of MBHs at $z=6$ that power quasars with bolometric luminosity larger than $10^{45}$ erg/s. Blue points: mass function derived by Willott et al. 2010 from a sample of $z=6$ quasars.}
\label{MF6}
\end{figure}

\section{Accretion and Spin evolution: results}
\begin{figure}
\includegraphics[width= \columnwidth]{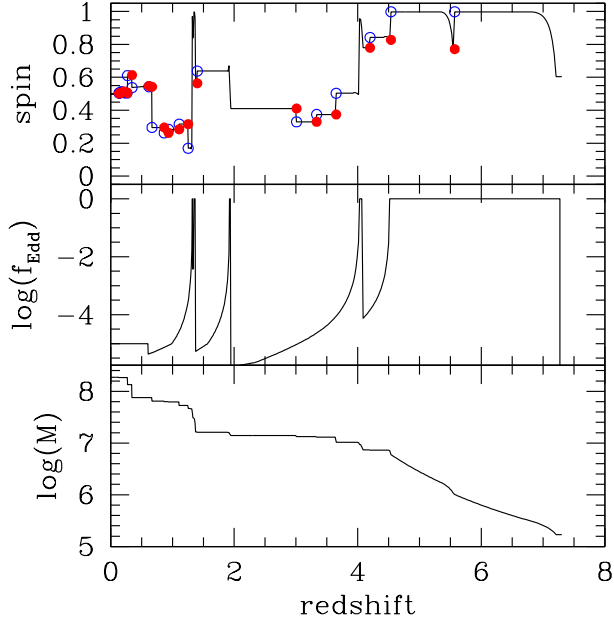}
\caption{Bottom panel: mass growth of a MBH in a galaxy that becomes a large spheroid by $z=0$. The evolution is extracted from a merger tree describing a dark matter halo of mass $4\times 10^{13} \msun$ at $z=0$. The  galaxy is the central galaxy in a group-sized halo.  Middle panel: evolution of the Eddington rate vs redshift. Top panel: evolution of the spin parameter. After an early phase of rapid accretion and growth the accretion rate declines and the MBH is fed by stellar winds only (quiescent phase) in the past $\simeq$ 6 Gyr. Today's spin is defined by a MBH-MBH merger, with a mass ratio $q=0.35$ that occurred $\sim$3 Gyr ago. Here the spin of the MBH before a MBH-MBH coalescence is shown as a blue empty circle, an the spin after coalescence as a red filled dot.}
\label{ell}
\end{figure}

\begin{figure}
\includegraphics[width= \columnwidth]{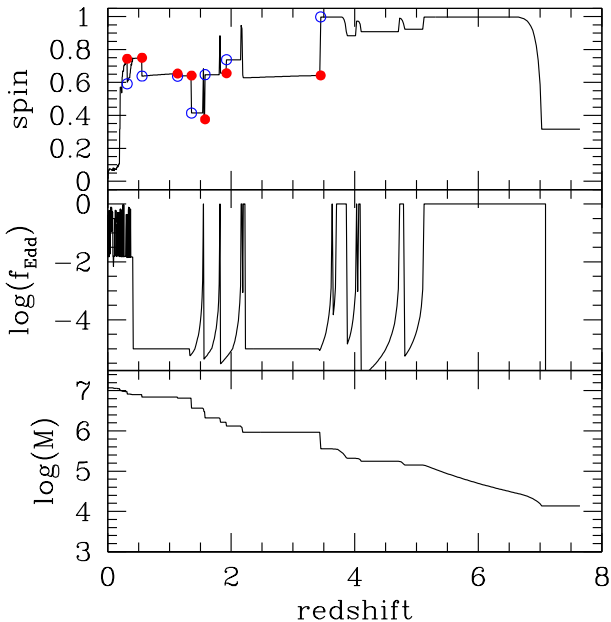}
\caption{Evolution of a MBH in a galaxy that becomes a Milky Way-type disk by $z=0$. The evolution is extracted from a merger tree describing a dark matter halo of mass $10^{12} \msun$ at $z=0$. The  galaxy is the central galaxy in a Local Group-sized halo.  Note the occurrence of MC accretion at $z<0.5$. Panels, lines and symbols as in Fig.~\ref{ell}.}
\label{disc_s}
\end{figure}

In Figure~\ref{ell} we show examples of  spin evolution of an MBH hosted in a large spheroid today along its cosmic history.  Most of the accreted MBH mass is accumulated during episodes of efficient growth at early times (up until $z\sim 2$). At lower redshift the MBH grows mostly through MBH-MBH mergers. While different prescription for ``chaotic" or ``coherent" post-feedback phases lead to different histories for the MBH spin, the final spin is set by a MBH merger between two roughly equal mass systems occurring at late cosmic times. This is noticeable as a small jump from $a=0.6$ to $a=0.5$ in the top panel of the figure at $z\simeq 0.3$.  
In the case of disk galaxies, late phases of MC accretion at substantial accretion rates ($>10^{-3}$ in Eddington units ) contribute to setting the final spin of the MBH (Fig.~\ref{disc_s}). 

\begin{figure}
\includegraphics[width= \columnwidth]{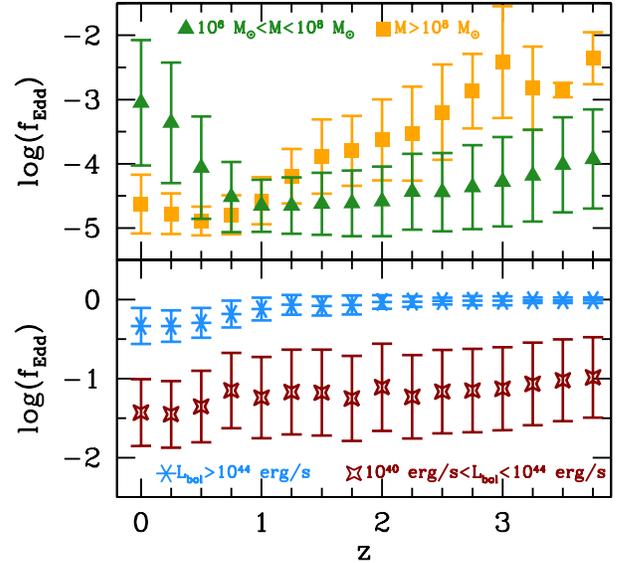}
\caption{Mean values of the  Eddington ratio (logarithmic units) as a function of redshift. The top panel shows a sample selected on the MBH mass ($10^6\msun<M<10^8 \msun$ and $M>10^8 \msun$) showing classic ``anti-hierarchical" behavior. The bottom panel focuses instead on luminosity-selected AGN, for which the typical accretion rate is much higher.}
\label{fedd}
\end{figure}

Statistically, Fig.~\ref{fedd} shows the evolution of the logarithmic Eddington ratio as a function of redshift. The top panel shows a sample selected on the MBH mass ($M>10^6 \msun$ and $M>10^8 \msun$) showing classic ``anti-hierarchical" behavior (Merloni 2004), with the the most massive MBHs being more active at earlier cosmic times. At a given luminosity threshold the typical Eddington ratio decreases slightly at late times (bottom panel of Fig.~\ref{fedd}), tracking instead the overall increase in the mass of MBHs. 

%


\begin{figure}
\includegraphics[width= \columnwidth]{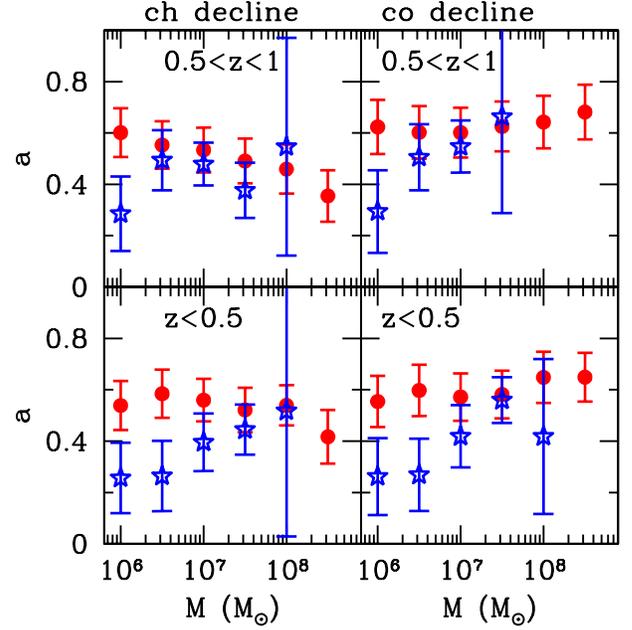}
\caption{Spins in galaxies of different morphologies as a function of MBH mass (mean and 1-$\sigma$ dispersion). Filled circles: gas-poor galaxies (spheroids). Stars: gas-rich galaxies (disks).  Bottom:  $z<0.5$. Top: $0.5<z<1$. Lower mass MBHs in gas-rich galaxies tend to spin less rapidly than higher mass ones, and also less rapidly than MBHs in gas-poor galaxies. }
\label{disc_sph} 
\end{figure}

In Fig.~\ref{disc_sph} we compare the spins of MBHs hosted in disks and spheroids, in different redshifts bins. At low MBH mass ($M\sim10^6$ \msun)  MBHs hosted in gas-rich galaxies tend to have low spins. The spin distribution in gas-rich galaxies  tends to move towards higher spins as mass increases. This is mostly related to MC accretion. For the most massive MBHs most of the growth occurs earlier on through merger-driven accretion that tends to spin-up MBHs. Accretion of MCs does not modify much the spins of these MBHs because the total angular momentum accreted through MCs is less than the total angular momentum the MBH has (i.e., the total mass accreted by the MBH through MCs is much less than the mass of the MBH).  On the other hand, for low-mass MBHs the mass accreted in MCs is of the same order as the MBH mass, therefore they have a stronger effect on the spin distribution, lowering the typical spin of low-mass MBHs (we remind that we have assumed that MCs accrete isotropically on MBHs).  The distribution of spins of MBHs hosted in gas-poor galaxies has little dependence on mass and redshift. In these galaxies, in general, most MBHs have spin $a\sim 0.4-0.8$. Spins tend to slightly decrease as MBH mass increases.  We find no strong dependence on  whether accretion occurs mostly chaotically or coherently  after `feedback' effects take place, except at the highest masses. We have run a test case where we have artificially ``turned-off" spin evolution via MBH-MBH mergers (while keeping the mass increase through mergers). In general, the effect of MBH-MBH mergers is to decrease the spins of the most massive MBHs in the case of coherent post-feedback phase, while it increases their spins in the case of chaotic post-feedback phase.

\begin{figure}
\includegraphics[width= \columnwidth]{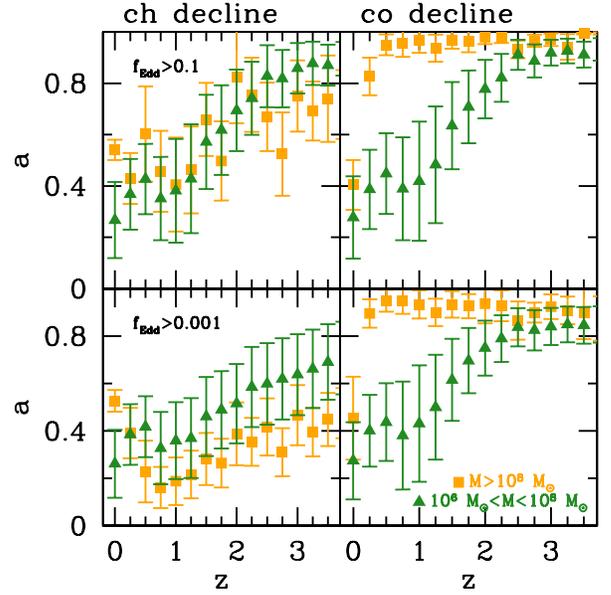}
\caption{Spins in AGN, selected by mass and Eddington ratio.  Triangles: MBH mass $10^6 \msun<M<10^8 \msun$. Squares: MBH mass $>10^8 \msun$. Top: QSOs accreting at high accretion rate (in Eddington units) $f_{\rm Edd}>0.1$. Bottom: AGN accreting at all accretion rates $f_{\rm Edd}>0.001$. Left:  chaotic accretion during the decline of the quasar phase. Right:  coherent accretion during the decline of the quasar phase. The signature of feedback on the angular momentum of nuclear gas is strongest on high mass MBHs (compare right and left panels). In general AGN spins tend to decrease at late cosmic times.}
\label{QSO_AGN} 
\end{figure}

In Fig.~\ref{QSO_AGN} we focus on active MBHs. MBHs accreting at high rates, $f_{\rm Edd}>0.1$, have very large spins at all $z>2$. These are for the most part MBHs in the ``quasar" phase. At $1<z<2$ more systems are caught in the decline phase, and here, especially at high MBH masses, it becomes crucial whether accretion occurs mostly chaotically or coherently after the quasar phase (cf. left and right columns). This mass ($>10^8 \msun$) and reshift ($0.5<z<1$) range is the most suitable to probe how feedback affects the angular momentum of nuclear gas.  For low-mass BHs the spin distribution is mostly insensitive to the chaotic and coherent models (green triangles in Fig.\ref{QSO_AGN}), while there is a stronger impact on high-mass holes (orange squares in Fig.\ref{QSO_AGN}). Therefore the changes most affect the high-luminosity end of the luminosity function. Finally, at $z<0.5$ differences tend to disappear, as MC accretion becomes the dominant feeding mechanism. The behavior of lower accretion rate systems is similar, although more and more systems are in the ``decline" phase, and lower spins become more common if accretion occurs chaotically during this phase.  

\begin{figure}
\includegraphics[width= \columnwidth]{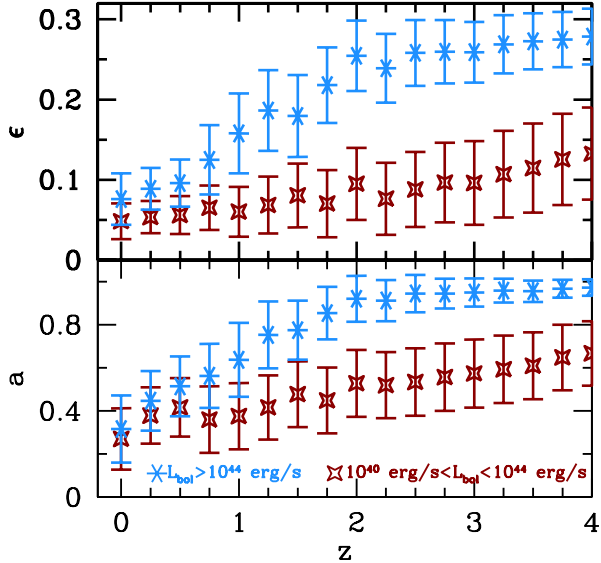}
\caption{Top: radiative efficiency of AGN  of different luminosities. Diamonds:  $10^{40}$ erg~s$^{-1}<L_{\rm bol}<10^{44}$ erg~s$^{-1}$. Asterisks: $L_{\rm bol}>10^{44}$ erg~s$^{-1}$. Bottom: spins of the same AGN. 30-40\% of low luminosity sources ($L_{\rm bol}\sim10^{40}$ erg~s$^{-1}$) are genuinely inefficient accretors. Sources with $L_{\rm bol}>10^{44}$ erg~s$^{-1}$ are powered by efficient accretors, and spin solely determines their radiative efficiency. }
\label{eps} 
\end{figure}

The spin distribution, however, does not necessarily map the radiative efficiency distribution (Fig.~\ref{eps}, also note that the mean efficiency differs, mathematically, from the efficiency corresponding to the mean spin), if a large population of sources have geometrically thick and optically thin  accretion disks where the radiative efficiency may be suppressed with respect to the mass-to-energy conversion efficiency.  Especially at low luminosity  and low redshift, many AGN are radiatively inefficient sources and the MBH spin is not relevant in determining their radiative efficiency. At low luminosity and high redshift AGN are a mixture of low mass MBHs accreting at high rates in radiatively efficient fashion, and higher mass MBHs accreting at low rates,  while at low redshift  most low luminosity sources are genuinely inefficient accretors. At high luminosity, instead, the signature of spin evolution with redshift is evident.  Fig.~\ref{QSO_AGN}  and Fig.~\ref{eps} show that our model is in very good agreement with the evolution of radiative efficiency with redshift  derived observationally by Li et al. (2012, see their Fig.~7), and by the suggestion of Li et al. and Shankar et al. (2011) that radiative efficiency may increase with black hole mass. 

\begin{figure}
\includegraphics[width= \columnwidth]{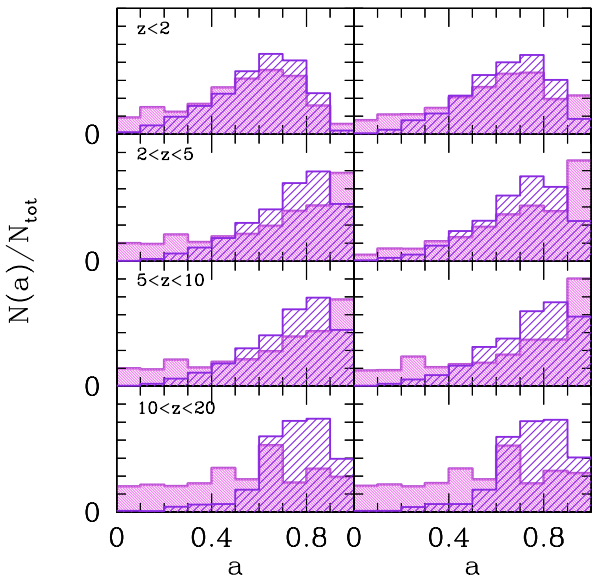}
\caption{Spins of the primary MBHs in a binary prior to coalescence (pink filled histogram) and spin of the newly merged MBH post  coalescence (violet hatched histogram) . Left:  chaotic accretion during the decline of the quasar phase. Right:  coherent accretion during the decline of the quasar phase.}
\label{mergers} 
\end{figure}

Finally, in Fig.~\ref{mergers} we show the distribution of MBH spins before and after a MBH-MBH coalescence. As expected, if MBHs have large spins pre-coalescence, MBH mergers tend to spin down the systems, and viceversa if MBHs have low spins pre-coalescence, MBH mergers tend to increase spins.  We here show  the spin of the primary MBHs in a binary prior to coalescence, as this is the quantity that gravitational wave observatories such as eLISA can measure.


%

\section{Discussion and Conclusions}
We  developed a model for the evolution of MBHs that takes into account several physical mechanisms of MBH growth: MBH-MBH mergers, merger-driven accretion, stochastic accretion, and accretion of recycled gas. This model, however, does not include MBH feeding through disk instabilities, nor the burst phase of recycled gas feeding in elliptical galaxies.  Under a series of plausible assumptions we have derived the growth of MBHs, the properties of the AGN population and the evolution of MBH spins. Our approach produces a population of MBHs and AGN consistent with the observed one in terms of, e.g., luminosity function of AGN, relationship between MBHs and their hosts, high-redshift quasars. 

The main results of our models of  MBH  evolution  can be summarized as follows:

\noindent $\bullet$  At high-redshift MBHs grow mostly by merger-driven accretion, while at later times other  channels become more important. In gas-rich galaxies, MC accretion dominates the growth of low-mass  ($<10^7 \msun$) MBHs at $z<2$. In gas-poor galaxies, MBH-MBH mergers are the main growth channel, especially at high MBH mass ($>10^7 \msun$);

\noindent $\bullet$ The mass of most active black holes decreases with increasing cosmic time, in ``anti-hierarchical" fashion. Sustained accretion grows the most massive black holes since early times without overproducing the MBH population as a whole;

\noindent $\bullet$  MBH spins tend to be  larger at redshifts $z > 2$, typically $a \ge 0.8$. They result from massive, coherent accretion events triggered by  major mergers. This result is in general agreement with the trend found by Barausse (2012) using a complementary, more refined approach to modeling galaxy evolution;

\noindent  $\bullet$   A significant drop in the average value of MBH  spins takes place at $z<2$.  This is caused by the increasing number of dry MBH-MBH mergers at lower redshifts in the case of spheroids. Additionally, a dramatic drop is predicted at $z < 0.5$, for low-mass MBHs in gas-rich galaxies. This is due to low-mass, chaotic accretion events involving capture of  molecular clouds;

\noindent  $\bullet$ In general, in gas-rich galaxies at $z<1$ low-mass MBHs tend to spin slightly less rapidly than high-mass MBHs. The difference is less pronounced in gas-poor galaxies. As MBH mass increases the distributions become more similar. At the highest masses ($>10^8 \msun$) the statistics are poor in the case of disk galaxies (between a couple and $\sim$50 objects per bin), as the most massive among MBHs tend to reside in gas-poor galaxies.  

\noindent $\bullet$  If outflows do not affect the angular momentum of nuclear disks,  and accretion proceeds  coherently both in the quasar and decline phase, the spin distribution and its evolution is not very differed for highly accreting vs low accreting MBHs. Differences are clearer in the population of the most massive black holes ($>10^8 \msun$).

\noindent $\bullet$  If quasar feedback disrupts the nuclear disk feeding the MBH and accretion proceeds chaotically in the decline phase, the spin distribution and its evolution shows a stronger dependence on mass,  but not on morphology, again, except at the highest masses ($>10^8 \msun$) and $z<1$. The same comment on statistical significance as above applies here as well.

Qualitatively similar results have been obtained  by  Li, Wang \& Ho (2012) on  observational grounds. They inferred  the MBH  spin evolution by tracing the evolution  of the radiative efficiency of accretion flows, using the continuity equation  for the MBH number density. Both our and Li et al. results seem to contradict  the predictions of the `spin paradigm' scenario according to which the jet  production efficiency - and therefore, radio-loudness of AGN - should reflect the MBH spin distribution and its evolution (Wilson \& Colbert 1995; Hughes \& Blandford 2003). Applying such a scenario to our results,  one should expect the radio-loud fraction of AGN to be much larger at high redshifts than in the present epoch. At least in the case of quasars,  an opposite trend has been inferred, i.e. such a fraction has been suggested to decrease with  redshift (Jiang et al. 2007), although Volonteri et al. (2011c) find that the radio-loud fraction is roughly constant with redshift for the most luminous sources ($L>10^{47}$ erg/s). For high-redshift blazars powered by $M>10^8 \msun$ MBHs, also, activity seems to peak around $z\sim 4$ \cite{VHG}. 
Based on these indications, it may be that spin plays a more important role in powering jets at high accretion rates. Intermittent jet production in high accretion-rate AGN  (Sikora et al. 2007), may occur in similarity to that observed in the galactic  micro-quasar GRS 1915+105 (Fender et al. 2004).  However, this does not explain the finding that radio-loud quasars reside in more massive and denser environments than the radio-quiet ones (Shen et al. 2009; Donoso et al. 2010).  Regarding low-accretion rate AGN, given the similar  MBH spin distributions in the `disk' and `spheroid' sub-populations  no double upper-bound pattern is expected  to emerge in the radio-loudness vs. Eddington-ratio plots, in contrast  to that found  by Sikora et al. (2007, but see Broderick and Fender 2011 for a discussion on how large-scale environmental effects may affect  low-frequency, low resolution radio loudness measurements, causing a larger dispersion in observed radio loudness). 

It should be noted, however, that  all these contradictions do not  necessary jeopardize idea of the dominant role of rotating BHs in powering  of extragalactic jets. This is because the Blandford-Znajek mechanism responsible  for the extraction of rotational BH energy depends not only on the BH spin,  but also on the  magnetic flux threading the MBH. As Lubow et al. (1994) demonstrated  it  is rather impossible  to cumulate in  the center a sufficiently large  magnetic flux to power strong jets if the accretion  proceeds via geometrically thin disks. Hence the AGN that  become efficient jet producers are those that have passed through the super-Eddington,  advection dominated accretion phase at least at the beginning of the massive accretion event, or  those which were operating in the very sub-Eddington, advection  dominated accretion regime prior to  the massive accretion event  (Igumenshchev 2008; Cao 2011;   McKinney et al. 2012).  Therefore noting that radiogalaxies as well as radio-loud quasars reside on average in denser environments than radio-quiet AGN  (Wake et al. 2008; Mandelbaum et al. 2009; Lietzen et al. 2011), this might  imply that the broad range of AGN  radio-loudness  (see Sikora et al. 2007 and refs. therein) is mainly determined by the range of environmentally conditioned net magnetic fluxes collected in AGN centers (Sikora et al. 2012; Sikora \& Begelman 2014).

Spin plays a fundamental role in determining the mass-to-energy conversion efficiency, and as a consequence in timing the MBH cosmic evolution through the growth rate (Equation~1). There is a close connection between the spin distribution, the average radiative efficiency and  {So{\l}tan's} argument. Many authors have carried out direct tests of the {So{\l}tan's} argument, either using the Cosmic X-ray Background Radiation   to derive the total energy density released by
the accretion process (Fabian \& Iwasawa 1999), or by using the observed AGN luminosity functions (Yu \& Tremaine 2002; Marconi et al. 2004; Merloni \& Heinz 2008).   The radiative efficiencies needed to explain the relic population are within the range $\approx 0.06 \div  0.20$, with some tension among the published results that can be traced back to the particular choice of AGN LF and/or scaling relation assumed to derive the local mass density.

Recently, Gilfanov and Merloni (2013), have summarized  our current estimate of the (mass-weighted)  average radiative efficiency in just one formula,  relating $\langle \epsilon \rangle$ to various sources of systematic errors in the determination of supermassive black hole mass density:

\begin{equation}
1-\xi_i-\xi_{\rm CT}+\xi_{\rm lost}=\frac{1-\langle \epsilon \rangle}{\langle \epsilon \rangle}R
\label{eq:soltan}
\end{equation}
where $\xi_0=\rho_{\rm BH,z=0}/ 4.2\times 10^5 M_{\odot} {\rm Mpc}^{-3}$ is the local ($z=0$) MBH mass density in units of  4.2$\times 10^5 M_{\odot} {\rm Mpc}^{-3}$ (Marconi et al. 2004) and using the integrated bolometric luminosity function from Hopkins et al. (2007), they obtain$R\sim 0.075/\xi_0$. Here $\xi_i$ is the mass density of black holes at the highest redshift probed by the bolometric luminosity function, $z \approx 6$, in units of the local one, and encapsulates  uncertainties on the process of MBH formation;  $\xi_{\rm CT}$ is the fraction of SMBH mass density (relative to the local one) 
grown in unseen, heavily obscured,  Compton Thick AGN, still missing from our census;  finally, $\xi_{\rm lost}$ is the fraction black hole mass contained in ``wandering'' objects, that  have been ejected from a galaxy nucleus following, for example, a merging event and the subsequent production of gravitational wave, the net momentum of which could induce a  kick capable of ejecting the black hole form the host galaxy.  The model presented in this paper allows us to provide an estimate of two of the unknowns in Equation~\ref{eq:soltan}: $\xi_i$ and $\xi_{\rm lost}$. In our models they are both of order 0.1 and they roughly cancel each other.  We can also directly estimate $\langle \epsilon \rangle=(\sum \epsilon_{i} \Delta M_{MV,i})/\sum\Delta M_{MV,i}=0.13-0.18$ (the sum is done for all accreting MBHs starting from $z=20$ down to $z=0$, so it is an average over mass and time), leading to an estimate of $\xi_{\rm CT}\sim 0.5$.  Current estimates of the Compton Thick AGN fraction (that are, however, not ``mass weighted" as in the formalism of Equation~\ref{eq:soltan}, see also the note in the Appendix) based on local Universe and models of the X-ray background range between 20 and 50\% (e.g., Akylas et al. 2012).

\acknowledgments
We  thank M. Dotti for helpful discussions and comments. MV acknowledges funding support from NASA, through award ATP NNX10AC84G; from SAO, through award TM1-12007X, from NSF, through award AST 1107675, and from a Marie Curie Career Integration grant (PCIG10-GA-2011-303609). JPL was supported in part by a grant from the French Space Agency CNES,  by the Polish Ministry of Science and Higher Education within the project N N203 380336, and by the Polish National Science Centre through grant DEC-2012/04/A/ST9/00083.

\appendix
\section{Black hole growth}
We here recall how one can describe the growth of a MBH as a function of a constant or variable Eddington rate. We start by defining $f_{\rm Edd}=L/L_{\rm Edd}$, and $L_{\rm Edd}=Mc^2/t_{\rm Edd}$, where $t_{\rm Edd}=\frac{\sigma_T \,c}{4\pi \,G\,m_p}= 0.45$ Gyr  and $f_{\rm Edd}$ represents the Eddington fraction. Therefore, if the accretion rate is $\dot{M}_{in}$, and $\dot{M}$ is the mass that goes into increasing the MBH mass:
\beq
L=\epsilon \, \dot{M}_{in} c^2= f_{\rm Edd} L_{\rm Edd} c^2
\eeq
and $dM = (1-\epsilon) dM_{in}$ and  
\beq
\frac{dM}{dt} = \frac{1-\epsilon}{\epsilon} L_{\rm Edd} f_{\rm Edd} c^2=\frac{1-\epsilon}{\epsilon} f_{\rm Edd} \frac{M\, c^2}{t_{\rm Edd}}, 
\eeq
one obtains 
\beq
\frac{dM} {M}  =  \frac{1-\epsilon}{\epsilon} f_{\rm Edd}\, t_{\rm Edd}^{-1} dt.
\eeq  

If $\epsilon$ and $f_{\rm Edd}$ are constant over the time of integration, then:
\beq
\int  M^{-1} dM  = \int \frac{1-\epsilon}{\epsilon} f_{\rm Edd}\, t_{\rm Edd}^{-1} dt,
\eeq  
and the MBH mass grows as
\beq
M(t+\Delta t)=M(t)  \exp\left({f_{\rm Edd}\frac{\Delta t}{t_{\rm
      Edd}}\frac{1-\epsilon(t)}{\epsilon(t)}}\right),
\eeq
while if, for instance $f_{\rm Edd}$ is a function of time, one has to self-consistently integrate:
\beq
\int  M^{-1} dM  = \int \frac{1-\epsilon}{\epsilon}  t_{\rm Edd}^{-1}  f_{\rm Edd}(t) \,dt,
\eeq  
as shown in section 4.2.

Note that this mathematical formalism differs from the approximate form $M_{BH}=(1-\epsilon) \dot{M}_{in}  \Delta t$ (the two expressions agree in the limit $\frac{1-\epsilon}{\epsilon} f_{\rm Edd}\, \Delta t \, t_{\rm Edd}^{-1} \rightarrow 0$), and the  mass-to-energy conversion efficiency appears within an exponential. This causes some inconsistency with the  ``standard" formalism used to evaluate the  mass-to-energy conversion efficiency in {So{\l}tan's} argument that adopts the simplified expression. At fixed $\epsilon$, $f_{\rm Edd}$ and $\Delta t$ the approximate expression underestimates the mass growth, and therefore, statistically,  {So{\l}tan's} argument tends to underestimate $\epsilon$ with respect to our formalism.

\section{Alignment of black hole spins in accretion discs}
In this paper we have assumed that most MBHs evolve in thin accretion discs where the importance of jets and magnetic fields is limited. In this case warp propagation occurs diffusively \citep{1975ApJ...195L..65B,1983MNRAS.202.1181P}.  In thick accretion discs ($H/R>\alpha$) warp propagation occurs instead  through bending waves \citep{2000MNRAS.315..570N}, while in magnetized discs with jets a ``magneto-spin alignment" mechanism has been recently discovered in numerical simulations \citep{McKinney}. 

We refer the reader to \cite{2000MNRAS.315..570N,2013ApJ...768..133S} and references therein for a full discussion of the mathematical treatment and the differences between diffusive and wave propagation, and we summarize here the relevant information. Bardeen \& Petterson (1975) showed that a viscous disc would be expected to relax to a form in which the inner regions become aligned with the equatorial plane of the black hole (Lense-Thirring precession) out to a transition radius, beyond which the disc remains aligned with the outer disc. This because the Lense-Thirring precession rate drops off sharply as the radius increases.  The transition radius ($r_{tr}$) is expected to occur approximately where the rate at which Lense-Thirring precession is balanced by the rate at which warps are diffused or propagated away. 

In the diffusive regime, the warping of the disc is counteracted by diffusion of the warp, which acts over the diffusion timescale $t_{\rm diff}\sim 4 r^2\alpha/(H^2\Omega)$. In the bending wave regime warps evolve on the sound crossing time $t_{\rm wave}\sim r/c_s$. The critical variable to determine the timescales over which (anti)alignment occurs is therefore the transition radius between the inner equatorial disc and the outer tilted disc.   \cite{2000MNRAS.315..570N} perform a numerical parametric study of both regimes and they conclude that, although the processes differ, the typical (anti)alignment timescale is well described by the formalism introduced by \cite{1978Natur.275..516R} and subsequently studied in greater detail by \cite{1996MNRAS.282..291S} and \cite{NatarajanPringle1998}. Similar conclusions are recently obtained by \cite{2013ApJ...768..133S}. We note, however, that \cite{2005ApJ...623..347F}, \cite{2007ApJ...668..417F}, and \cite{2011ApJ...730...36D} find no alignment in their three-dimensional general relativistic magnetohydrodynamic simulations of tilted discs. After Lense-Thirring precession causes the disk to warp, the propagation of the warp stops at the radius in the disk where the sound-crossing time becomes shorter than the precession time, and the disc remains tilted.

Finally, we remark that \cite{McKinney}, also using three-dimensional general relativistic magnetohydrodynamical simulations, recently proposed that in magnetized disks the frame-dragging forces cause first the MBH magnetosphere to align with the MBH spin axis, and then the disc loses the misaligned component of its angular momentum and reorients with the magnetosphere, at least at small radii (the outer disc may remain tilted, see their Table 2).

\end{document}